\documentstyle[itzidef]{mn}

\def\DIRFIGS{}
\def\kelvin{\thinspace\hbox{K}}
\def\percucm{\thinspace\hbox{$\hbox{cm}^{-3}$}}
\def\ergsec{\thinspace\hbox{$\hbox{erg}\thinspace\hbox{s}^{-1}$}}

\input epsf

\title[SN~1988Z: Spectro-photometric catalogue and energy estimates]
{SN~1988Z: Spectro-photometric catalogue and energy 
estimates\thanks{Based on observation carried out at ESO La Silla}
}
\author[I. Aretxaga, S. Benetti, R.J. Terlevich, A.C. Fabian, 
E. Cappellaro, M. Turatto and M. Della Valle]{
Itziar Aretxaga$^{1,2}$\thanks{Present address: INAOE, Puebla, Mexico}, 
S. Benetti$^3$, R. J. Terlevich$^4$\thanks{Visiting Professor at 
INAOE, Puebla, Mexico}, A.C. Fabian$^4$, \\ \\
{\LARGE E. Cappellaro$^5$, M. Turatto$^5$ and M. Della Valle$^{6,7}$} \\
$^1$ Max-Planck-Institut f\"ur Astrophysik, Karl Schwarzschildstr. 1,
Postfach 1523, 85740 Garching, Germany\\
$^2$ Instituto Nacional de  Astrof\'{\i}sica, \'Optica y Electr\'onica, 
Aptdo. Postal 51 y 216, Puebla, Mexico\\
$^3$ Telescopio Nazionale
Galileo, Aptdo. Correos 565, E-38700 Santa Cruz de La Palma,
Canary Islands, Spain\\
$^4$ Institute of Astronomy, Madingley Road, Cambridge CB3 0HA, U.K.\\
$^5$ Oss. Astronomico di Padova, vicollo dell'Osservatorio
5, I-35122, Italy\\
$^6$ Dipar. di Astronomia, Univ. di Padova, vicollo dell'Osservatorio
5, I-35122, Italy\\
$^7$ European Southern Observatory, Alonso de Cordova 3107,
Vitacura, Casilla 19001, Santiago 19, Chile. 
}
\date{}

\begin{document}

\maketitle

\begin{abstract}

We present a spectro-photometric catalogue
of the evolution of supernova 1988Z which combines
new and published observations in the radio, optical and 
X-ray bands, with the aim of offering a comprehensive view of the
evolution of this object and deriving the total energy radiated 
since discovery. The major contribution to the total
radiated energy comes at optical  to X-ray frequencies, with
a total emission of at least 
$2 \times 10^{51}$~erg (for \Ho50) in 8.5 years. A model-dependent
extrapolation of this value indicates that the total radiated energy
may be as high as $10^{52}$~erg. 
The high value of the radiated energy supports a scenario in 
which most
of the kinetic energy of the ejecta is thermalized 
and radiated in a short
interaction with a dense circumstellar medium of nearly constant density. 
In this sense, 1988Z is not a supernova 
but a young and compact supernova remnant.

\end{abstract}



\section{Introduction}

The explosion of massive stars, thought to take
place as the result of pair formation, collapse of a Fe core or
neutronization of an ONeMg degenerate core, leads to supernovae (SN) 
that show H signatures in their spectra and thus 
are classified as type II.
These objects display a wide range of observational properties
which justifies the existence of the two photometric sub-classes,
II-P and II-L (Barbon, Ciatti \& Rosino 1979),
and the recent introduction of the peculiar spectroscopic 
sub-group IIn (Schlegel 1990).


The spectra of SN~IIn are characterized by the 
presence of prominent narrow emission lines sitting on top of broad
components of FWHM$\approx 15000$~km/s at maximum light 
which look very similar to
those of Seyfert~1 nuclei and QSOs (Filippenko 1989). They 
don't show the characteristic broad P-Cygni signatures of standard SN,  
although narrow P-Cygni profiles are detected 
in some cases at high spectral resolution (SN~1997ab,
Salamanca et al. 1998; SN~1995G, ESO data archive;
SN~1997eg, Tenorio-Tagle G., priv. communication). SN~IIn 
are normally associated with regions of recent star formation.
Despite these general characteristics, SN~IIn as a group exhibit some
heterogeneity (see Filippenko 1997, Turatto et al. 1999).

  SN~1988Z, one of the most extensively observed SN of this class,
was discovered in 1988 December 12 in the galaxy
M+03-28-022 (Zw 095-049) 
(Cappellaro \& Turatto 1988, Pollas 1988a). 
Earlier observations in late spring 1988 
show no evidence of the event (Turatto et al. 1993a).
This  is an exceptionally bright and peculiar SN in its
spectro-photometric
properties:
\begin{description}
\item[$\bullet$] It is characterized by an extremely slow decay of
luminosity 
after maximum light, which makes it at day 600 approximately 5~mag
brighter in the $V$-band than standard SN~II-P or SN~II-L 
(Stathakis \& Sadler 1991).
\item[$\bullet$] It has a strong  \Ha\ emission, with 
peak luminosities of about $4 \times 10^{41}$~erg~s$^{-1}$ 
(for \Ho50)
at day 200   (Turatto et al. 1993a). This
prodigious luminosity is 5 orders of magnitude larger than that
of SN~1987A.
\item [$\bullet$] Very high-ionization coronal lines (e.g. [Fe
X]\ldo{6375},
[Fe XI]\ldo{7889-7892})
are identified in the optical spectra, at least until day 492 (Turatto et
al. 1993a).
\item [$\bullet$] At 2 to 20cm  it is one of the most powerful 
radio-SN in the sky,
with peak-luminosities up to 3000 times that of remnants like Cas~A
(Van Dyk et al. 1993).
\item[$\bullet$] 
Even 6~yr after maximum the SN shows an X-ray
emission of more than $10^{41}$~erg~s$^{-1}$ (Fabian \& Terlevich 1996).
\end{description}
The summary of these points is that this is one of the most energetic
radiative events known to be generated in a single stellar object.

These radiative properties have been interpreted in the light of 
quick re-processing of the mechanical energy of a SN explosion
by a dense circumstellar medium (CSM) (Chugai 1991, Terlevich et al. 1992,
Terlevich 1994, Chugai \& Danziger 1994, Plewa 1995).
As explained in the appendix,  radiative cooling is expected to become
important well before the thermalization of the ejecta is complete.
As a result, the shocked material undergoes a rapid condensation
behind both the leading and reverse shocks. These high-density thin
shells, the freely expanding ejecta and the still unperturbed interstellar
gas
are all ionized by the radiation produced in the shocks, and are
responsible for the complex emission line structure observed in 
these objects.

In the case of SN~1988Z, a direct measurement
of the CSM electron density was possible.
The value determined from 
the [OIII]\ldo{4363}\ to [OIII]\ldo{5007}\
forbidden line ratio in the early stages of the evolution
is between  $4 \times 10^6$ and
$1.6 \times 10^7$~\uniden\ (Stathakis \& Sadler 1991).
This high density points 
towards SN~1988Z in particular, and SN~IIn in general,
being young compact SN remnants (cSNR).

In this paper we present new optical and X-ray spectro-photometric
observations of the late evolution (4.3 to 8.5 yr) 
of this object, which combined with already published information on its
early evolution, aims at deriving the total energy radiated 
since discovery, and shedding new light 
into the mechanism
that generates such a long-lasting bright event.


\ifoldfss
  \section{Data set}
\else
  \section[]{Data set}
\fi

\subsection{Optical observations}
\vskip 1cm

\begin{table*}
 \begin{minipage}{140mm}
\begin{center}
\caption{Broad-band photometry}\label{obs_tab}
\begin{tabular}{lccccl}
\hline
    date     &  age         &        B                &    V
&        R              &   instrument   \\
  & days & mag  & mag & mag & \\ 
\hline
1993 Mar 27    & 1565 & $22.23\pm 0.26$& $22.08\pm 0.25$&$20.79\pm 0.20$&
NTT + EMMI \\
1994 Jan 14    & 1856 &                           &
&$21.36\pm 0.20$&3.6m + EFOSC1         \\
1994 Dec 30  & 2209 &                           &
&$21.71\pm 0.20$&NTT + EMMI \\
1995 Jan 10    & 2220 &                           & $22.73\pm
0.20$ &$21.70\pm 0.15$&NTT + SUSI \\
1996 Feb 14    & 2620 &                           &
&$21.87\pm 0.20$&3.6m + EFOSC1    \\
1997 Feb 10    & 2982 &                           &
& $22.04\pm 0.20$&3.6m + EFOSC1  \\
\hline
\end{tabular}
\end{center}
\end{minipage}
\end{table*}

\begin{table*}
 \begin{minipage}{140mm}
\begin{center}
\caption{Journal of spectroscopic observations} \label{spec_tab}
\begin{tabular}{lclccr}
\hline
     date     & age  & instrument  &  exp. &   range   & resol. \\
                 & days  &                    & min &   \AA   & \AA   \\
\hline
1993 Mar 27     & 1565    &  NTT + EMMI       &105 & 3750-9700 & 13  \\
1994 Jan 13-14     & 1859    &     3.6m + EFOSC1           &115 &
3700-6900 & 16  \\
1994 Dec 30   & 2209    &  NTT + EMMI       & 90 & 3700-8900 &  8   \\
1996 Feb 14     & 2620    &     3.6m + EFOSC1           &120&  3750-6950 &
16  \\
1997 Feb 9-10     & 2982    &     3.6m + EFOSC1           &240&  3730-6900
& 16  \\
\hline
\end{tabular}
\end{center}
\end{minipage}
\end{table*}

\begin{table*}
 \begin{minipage}{140mm}
\begin{center}
\caption{ROSAT HRI data for SN~1988Z. The 5th column tabulates the
probability that the counts seen are a random fluctuation in the
background. The flux in the 6th column has been corrected for Galactic
absorption.}
\begin{tabular}{@{}lcrrcr}
\hline
date & age & count rate & exp. & random chance & flux (0.2--2~keV)\\ 
     & days & $10^{-4}$~s$^{-1}$ & s & & $\times 10^{-14}$
erg~s$^{-1}$~cm$^{-2}$ \\	
\hline
1995 May 16--25 & 2355 & $11\pm4$  & 12287 & $2\times 10^{-5}$ &
$4\pm1.6$ \\
1996 Dec 14     & 2924 & $5\pm4$ &  6739 & 0.1 & $2\pm 1.6$ \\
1997 May 13--24 & 3085 & $3.7\pm 2$ & 34300 & 0.02 & $1.5\pm0.8$ \\ 
\hline
\end{tabular}
\end{center}
\end{minipage}
\end{table*}

\begin{figure*}
   \cidfig{5in}{20}{135}{585}{460}{\DIRFIGS 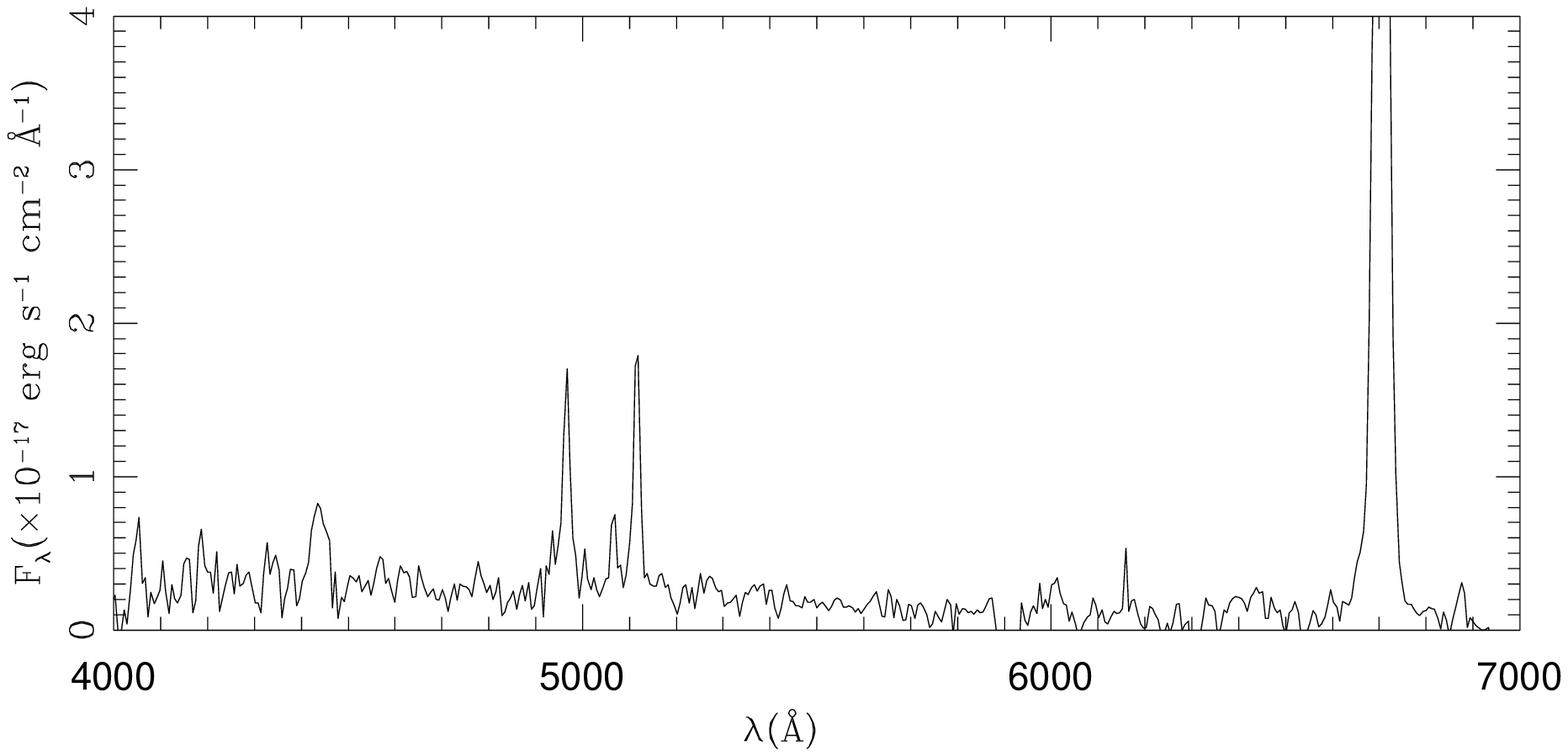}
   \caption{Spectrum of SN~1998Z on 1996 February 14.}
\end{figure*}

The new optical observations that we present in this paper
span a period of about 4 years, starting 4.3 years after 
discovery. 

$B$, $V$ and $R$-band photometry of SN~1988Z was obtained at the 
European Southern Observatory (ESO) in La Silla on
six nights between 1993 March 27 and 1997 February 10 
using the 3.6m telescope in combination with 
the ESO Faint Object Spectrograph and Camera  (EFOSC1) and
the 3.5m New Technology Telescope (NTT) with the ESO Multi Mode
Instrument (EMMI) or the Superb Seeing Imager (SUSI). 

When the nights were photometric
the magnitudes were calibrated using a number of Landolt's
(1992) standard stars. In
non-photometric nights the SN magnitudes were derived using the local
sequence defined in Turatto et al. (1993a). The SN magnitudes were
measured by fitting a point spread function with the {\sc ROMAFOT} package
of {\sc MIDAS}. 
This technique allows a good subtraction of the background and
gives reliable results even for faint objects (see Turatto et
al. 1993b). The journal of observations together with the
resulting magnitudes and an estimate of the internal
errors are reported in table \ref{obs_tab}, where age
refers to the time elapsed since the SN discovery, 1988 December 12 (JD
24447508). 

The journal of the spectroscopic observations is given in 
table \ref{spec_tab}, where for each spectrum column~1 gives 
the date of acquisition,
column 2 the age, column~3 
the instrumentation
used, column~4 the exposure time, column~5 the wavelength range 
and column~6 the resolution derived from the average FWHM of the night-sky
lines.
In two epochs we obtained spectra during two consecutive nights. In
these cases we co-added the spectra to reach higher
signal-to-noise ratios and included the cumulative exposure times in 
column 4.  The spectra were wavelength calibrated with He--Ar
arcs and flux calibrated using standard stars from the list of Hamuy
et al. (1994).  The absolute flux
calibration of the spectra was verified against the broad-band
photometry and the agreement was generally fair.

Figure~1 shows as an example of its late stages in evolution,
the spectrum of SN~1988Z on 1996 February 14. 
The earlier evolution of the spectrum can be seen in 
Stathakis \& Sadler (1991), Filippenko (1991) and Turatto et al. (1993a).

\subsection{X-ray data}

SN~1988Z has been observed three times with the {\it R\"ontgensatellit} 
(ROSAT) High Resolution
Imager (HRI), in 1995 (detection reported in Fabian \& Terlevich 1996), 
1996 and 1997. Table~3 summarizes these observations.

Only the first and
last observations provide a good detection of the object; the middle one
was too short. The count rate declined from $1.1\times 10^{-3}$ to
$3.7\times 10^{-4}$~s$^{-1}$ between 1995 May and 1997 May. Using Poisson
statistics on the numbers of counts detected, we find that the
probability that the average count rate 
was equal to or higher than that seen in
1995, and equal to or lower than that seen in 1997 is less than 0.2 per cent, 
adopting the weighted mean of both observations at both times.
This indicates that the decrease is statistically significant.

The 0.2--2~keV fluxes, corrected for absorption in our Galaxy
$N_H=2.5\times10^{20}$~cm$^{-2}$ (Burstein \& Heiles 1982), are given
in table 3, assuming a 1~keV bremsstrahlung spectrum. The 0.2--2~keV
luminosity therefore dropped from $(8\pm3.2)\times 10^{40}$~erg~s$^{-1}$
in 1995 to $(3\pm1.6)\times10^{40}$~erg~s$^{-1}$ in 1997.

These are probably underestimates of the X-ray luminosity.
The hydrogen column density can be much higher than that assumed
if the pre-ionized CSM is only partially ionized. 
Using 
the photoionization code 
CLOUDY 90 (Ferland 1997) 
and a description of the evolution of the leading shock of SN~1988Z
based on the model by Terlevich (1994) (see appendix) we estimate that
if the mass of the  CSM still to be swept is of the order of 1\Msun,
then the column density 
would be of the order of
$10^{22}$~cm$^{-2}$ at $t=12 t_{sg}$, where $t_{sg}$ 
refers to the 
time of the radiative onset of the remnant (eq. A3), 
around 250~days from core collapse for this object.
For a column density of $10^{22}$~cm$^{-2}$  
the previous estimates of the 
X-ray luminosity 
should be increased by a factor of
eight if we assume a 1~keV temperature.
The bolometric luminosities for the
assumed spectrum are 1.6 times greater.
We should emphasize that the absorption history is very dependent
in the total CSM mass.
For larger CSM masses the recombination
occurs earlier in the evolution and thus the soft X-ray correction is
larger.


ROSAT observations of SN~1986J, another SN~IIn, 
also show a high column density of 
$ 10^{22}$cm$^{-2}$ and a prominent X-ray luminosity
$2-3\times 10^{40}$~erg~s$^{-1}$ about 
9 years after the SN event (Bregman \& Pildis 
1992).

\subsection{Catalogue of radio, optical and X-ray photometry}

 \begin{table*}
 \begin{minipage}{140mm}
 \begin{center}
  \caption{Photometric evolution from X-ray to radio}
  \begin{tabular}{rcccccccccl}
\hline
Age & $f_{\mbox{\scriptsize 0.2 -- 2 keV}}$  &  $B$  &  $V$  &  $R$  & 
$f_{\Ha}$  &   $f_{\mbox{\scriptsize 2 cm}}$  &  
$f_{\mbox{\scriptsize 3.6 cm}}$   &  $f_{\mbox{\scriptsize 6 cm}}$ &
$f_{\mbox{\scriptsize 20 cm}}$  & reference \\
day & erg cm$^{-2}$ s$^{-1}$  &  mag  &  mag  &  mag 
& erg cm$^{-2}$ s$^{-1}$  &  mJy  &  mJy  &  mJy  &  mJy & \\
\hline
0       &     & 16.54 &       &       &      &      &      &      &      &
T93         \\
2      &      & 16.59 & 16.23 & 15.93 &      &      &      &      & & T93
\\
4      &      & 16.59 &       & 15.98 &      &      &      &      &	&
T93+IAUC   \\
23      &     & 16.71 & 16.46 & 16.04 &      &      &      &      &      &
SS91        \\
24      &     & 16.69 &       &       &      &      &      &      &      &
T93         \\
34      &     &       &       &       & 1478$\times 10^{-16}$  &      &
&      &      & SS91        \\
50      &     & 16.84 &       &       &      &      &      &      &      &
T93         \\
53      &     & 17.04 & 16.79 &       &      &      &      &      &      &
T93         \\
57      &     &       & 16.68 & 16.35 &      &      &      &      &      &
IAUC       \\
59      &     &       & 16.82 & 16.37 &      &      &      &      &      &
IAUC       \\
62      &     &       &       &       & 1669$\times 10^{-16}$  &      &
&      &      & SS91        \\
85      &     & 17.44 &       &       &      &      &      &      &      &
T93         \\
87      &     & 18.65 & 17.53 &       &      &      &      &      &      &
T93         \\
91     &     &       & 17.59 &       &      &      &      &      &      &
IAUC       \\
114     &     & 17.93 & 17.87 &       &      &      &      &      &      &
T93         \\
115     &     &       & 17.74 & 17.47 & 1449$\times 10^{-16}$  &      &
&      &      & T93         \\
117     &     &       & 17.91 &       &      &      &      &      &      &
T93         \\
137     &     & 18.03 & 17.74 & 17.53 &      &      &      &      &      &
T93         \\
199     &     &       &       &       & 1630$\times 10^{-16}$  &      &
&      &      & SS91        \\
282     &     &       & 19.02 & 17.80 &      &      &      &      &      &
T93         \\
392     &     &       &       &       &      &      & 0.67 &      &      &
vD93        \\
407     &     &       & 19.34 & 18.05 &      &      &      &      &      &
T93         \\
441     &     & 19.47 & 19.12 & 18.10 &      &      &      &      &      &
T93         \\
446     &     &       &       &       &      &      & 0.77 &      &      &
vD93        \\
451     &     &       & 19.19 &       &      &      &      &      &      &
T93         \\
464     &     &       &       &       & 1656$\times 10^{-16}$  &      &
&      &      & SS91        \\
474     &     &       &       & 18.09 &      &      &      &      &      &
T93         \\
492     &     &       &       &       & 1947$\times 10^{-16}$  &      &
&      &      & T93         \\
552     &     &       &       &       &      &      & 1.21 &      &      &
vD91        \\
557     &     &       & 19.54 & 18.38 &      &      &      &      &      &
SS91        \\
596     &     &       &       &       &      &      & 1.26 & 2.10 &      &
vD93        \\
653     &     &       &       &       &      &      & 1.38 &      & 1.15 &
vD93        \\
684     &     &       &       &       &  1019$\times 10^{-16}$  &      &
&      &      & T93         \\
717     &     &       &       &       &  936$\times 10^{-16}$  &      &
&      &      & T93         \\
731     &     & 20.78 & 20.16 &       &      &      &      &      &      &
T93         \\
736     &     &       & 20.25 &       &      &      &      &      &      &
T93         \\
738     &     &       &       &       &  730$\times 10^{-16}$  &      &
&      &      & T93         \\
743     &     & 20.79 & 20.09 & 18.88 &      &      &      &      &      &
T93         \\
751     &     &       &       &       &      &      & 1.78 & 2.09 & 1.43 &
vD93        \\
758     &     &       &       &       &  793$\times 10^{-16}$  &      &
&      &      & T93         \\
759     &     &       &       & 18.82 &      &      &      &      &      &
T93         \\
798     &     & 20.89 & 20.24 & 18.94 &  814$\times 10^{-16}$  &      &
&      &      & T93         \\
800     &     &       &       & 19.11 &      &      &      &      &      &
T93         \\
845     &     & 21.29 & 20.61 & 19.08 &      &      &      &      &      &
T93         \\
930     &     &       &       &       &      & 0.47 & 1.90 & 1.68 & 1.31 &
vD93        \\
1028    &     &       &       &       &      &      & 1.85 & 1.48 & 0.59 &
vD93        \\
1086    &     &       & 20.82 & 19.53 &      &      &      &      &      &
T93         \\
1095    &     &       & 20.59 & 19.48 &      &      &      &      &      &
T93         \\
1148    &     & 21.47 & 20.89 & 19.76&      &      &      &      &      &
T93         \\
1149    &     &       &       &       &  338$\times 10^{-16}$  &      &
&      &      & T93         \\
1160    &     &       &       &       &      & 1.05 & 1.67 & 1.50 & 1.17 &
vD93        \\
1420    &     &       &       &       &      & 0.91 & 1.57 & 1.22 & 0.91 &
vD93        \\
1487    &     &       &       &       &      & 1.45 & 1.72 & 1.04 & 0.78 &
vD93        \\
1565    &     & 22.02 & 21.92 & 20.67 &  124$\times 10^{-16}$    &      &
&      &	 & this paper   \\
1856    &     &       &       & 21.24 &  84.7$\times 10^{-16}$    &      &
&      &      & this paper   \\
2209    &     &       &       & 21.59 &  56.0$\times 10^{-16}$    &      &
&      &      & this paper   \\
2220    &     &       & 22.57 & 21.58 &      &      &      &      &      &
this paper   \\
2355    & 4$\times10^{-14}$\footnote{recalibrated value, was
 $3.5\times10^{-14}$ in FT96.} 
 &       &       &       &      &      &      &      &      & FT96
\\
2620    &     &       &       & 21.87 &  37.6$\times 10^{-16}$    &      &
&      &      & this paper   \\
2924    & 2$\times10^{-14}$ &       &       &       &      &      & &
&      & this paper        \\
2982    &     &       &       & 22.04 &  22.2$\times 10^{-16}$    &      &
&      &      & this paper   \\
3085    & 1.5$\times10^{-14}$ &       &       &       &      &      & &
&      & this paper        \\

\end{tabular}
\end{center}
\end{minipage}
\end{table*}

\begin{figure*}
   \cidfig{5in}{20}{144}{580}{710}{\DIRFIGS 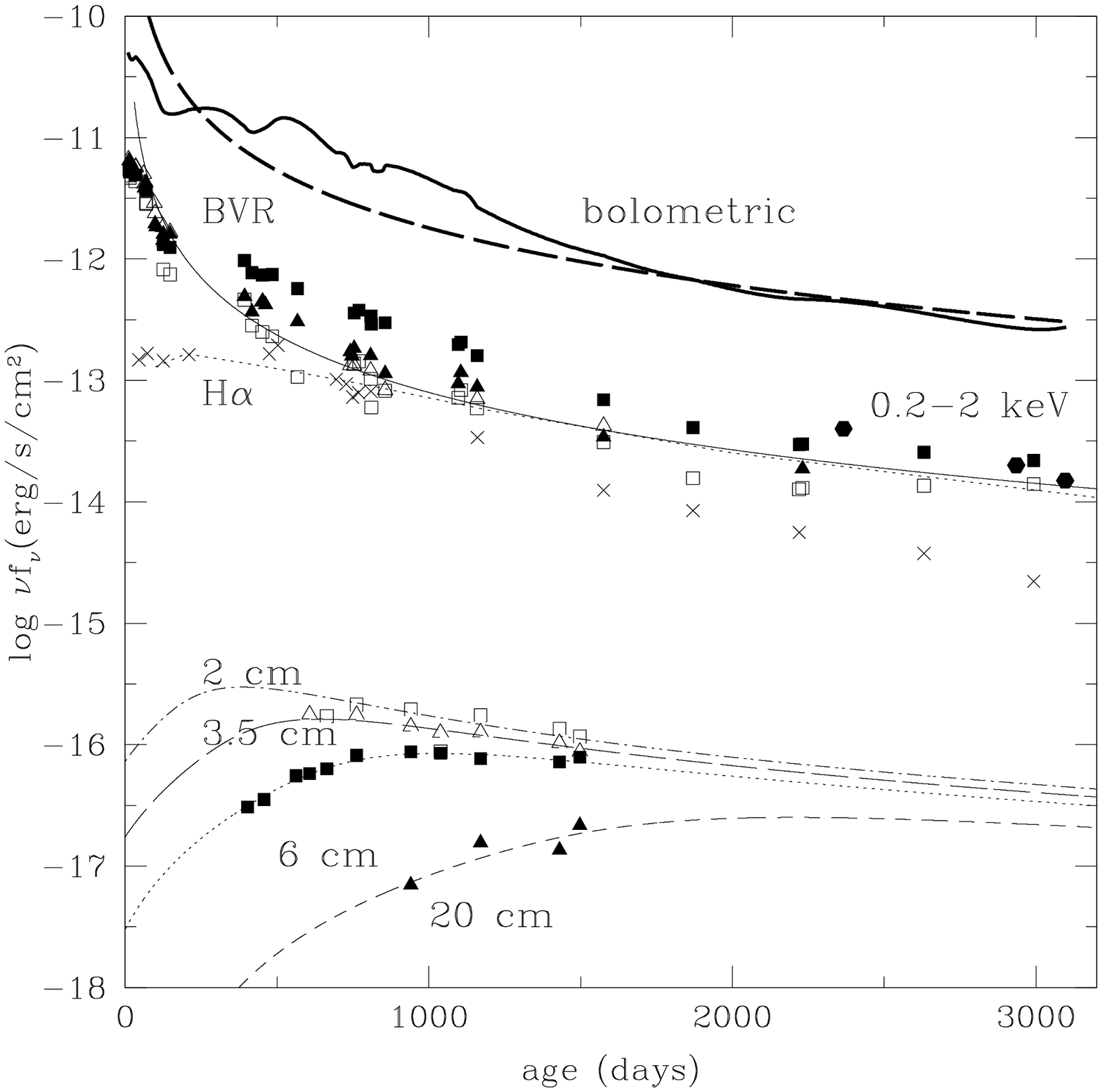}
   \caption{Evolution of the $\nu f_\nu$ light curve in radio to X-ray
bands. In the upper part of the panel open  triangles correspond to $B$-band,
solid triangles to $V$-band, solid 
squares to $R$-band, empty squares to $R_{\mbox{\scriptsize cont}}$ band,  
solid
hexagons to the ROSAT 0.2 to 2 keV band and crosses to \Ha. The thin solid and 
dashed lines are models for the bolometric light (scaled to the values of the
optical data) and \Ha\ evolution of a cSNR.
In the lower part of the panel, the VLA radio data at 2, 5.5, 6 and 20cm is 
represented with empty squares, empty triangles, solid squares and empty 
triangles. The line fittings correspond to the models of Van Dyk et al. 
(1993). The thick solid line on the top of the panel shows
the strong lower limit estimate while
the thick dashed line represents our best estimate 
of the bolometric light curve of SN~1988Z 
(see section 4.2).
}
\end{figure*}

A compilation of photometric data gathered from the literature,
combined with the new spectro-photometric observations introduced in
sections 2.1 and 2.2
is presented in table~4.
Column 1 gives the age since discovery.
Columns 2 to 10 give the flux and apparent magnitude of the object
in the 0.2 to 2 keV band from ROSAT; $B$, $V$, $R$-band photometry and
\Ha\ flux from spectroscopy obtained with ground-based
optical telescopes; and 2 to 20~cm fluxes from VLA. 
All fluxes are corrected for Galactic extinction using Seaton's law
(1979). Column~11
gives the reference from which the data was extracted, where 
SS91 refers to Stathakis \& Sadler (1991), T93 to Turatto et al. (1993a),  
FT96 to Fabian \& Terlevich (1996) and IAUC to the IAU circulars numbers
4691, 4696, 4742 and 4761. (Cappellaro \& Turatto 1988; 
Pollas 1988a, 1988b, 1989;
Gaskell \& Koratkar 1989)

Figure~2 shows a combined representation of the light curves of SN~1988Z
at different frequencies.
In the area occupied by the optical and X-ray data, solid crosses
correspond to \Ha, solid 
squares to $R$-band, solid triangles
to $V$-band, open  triangles to $B$-band and the solid
hexagons to the ROSAT 0.2 to 2 keV band.

The thin solid line that passes 
through the broad-band optical and X-ray points describes a $t^{-11/7}$
law. This is a semi-analytic approximation of 
the bolometric emission produced by a cSNR
when it evolves in a dense CSM of homogeneus density
(eq. A7) 
having  the canonical value 
of density of the CSM $10^7$~\uniden, and 
being scaled to match the $V$-band luminosity
of SN~1988Z. The overall decline is remarkably well followed by this law,
even at time-scales for which the assumptions of the semi-analytic 
approximations  are no
longer valid ($t\gsim250$~days).
The dashed line that passes through the \Ha\
points represents 
the \Ha\ luminosity predicted by the same models (Terlevich 1994),
with no intrinsic scaling. Although the models predict the peak
of \Ha\ luminosity correctly at day $\sim 200$, in the late
evolution of the SN they fail to reproduce the more pronounced decay 
by about an order of magnitude. 

The lines that pass through the radio points
are the models of an external thermal-absorbing gas combined with an
internal 
thermal-absorbing/non-thermal-emitting gas fitted by
Van Dyk et al. (1993) to their VLA observations.

Since \Ha, at the recession velocity of SN~1988Z
$v_r=6670$~\univel\ (Barbon, Cappellaro \& Turatto  1989),
is fully included in the $R$ pass-band 
we have estimated a \Ha\ line-free $R$-band magnitude subtracting an
interpolated \Ha\ flux from table~4 as
\begin{equation}
R_{\mbox{\scriptsize cont}}=
-2.5 \log \left( 10^{-R/2.5} - \frac{0.80 f(\Ha )}{f_R^0} \right)
\mbox{\ \ \  ,}
\end{equation}
where $f_R^0$ is the photometric zero of $R$-band in the same units as
f(\Ha), and the 0.80 factor represents the transmission value
of the redshifted \Ha\ line relative to the peak transmission value of
the $R$-band filter used in the observations (see figure~3).
The corrected $R$-band magnitudes ($R_{\mbox{\scriptsize cont}}$) are 
represented with empty squares in figure~2.

\begin{figure*}
   \cidfig{3in}{300}{425}{585}{710}{\DIRFIGS 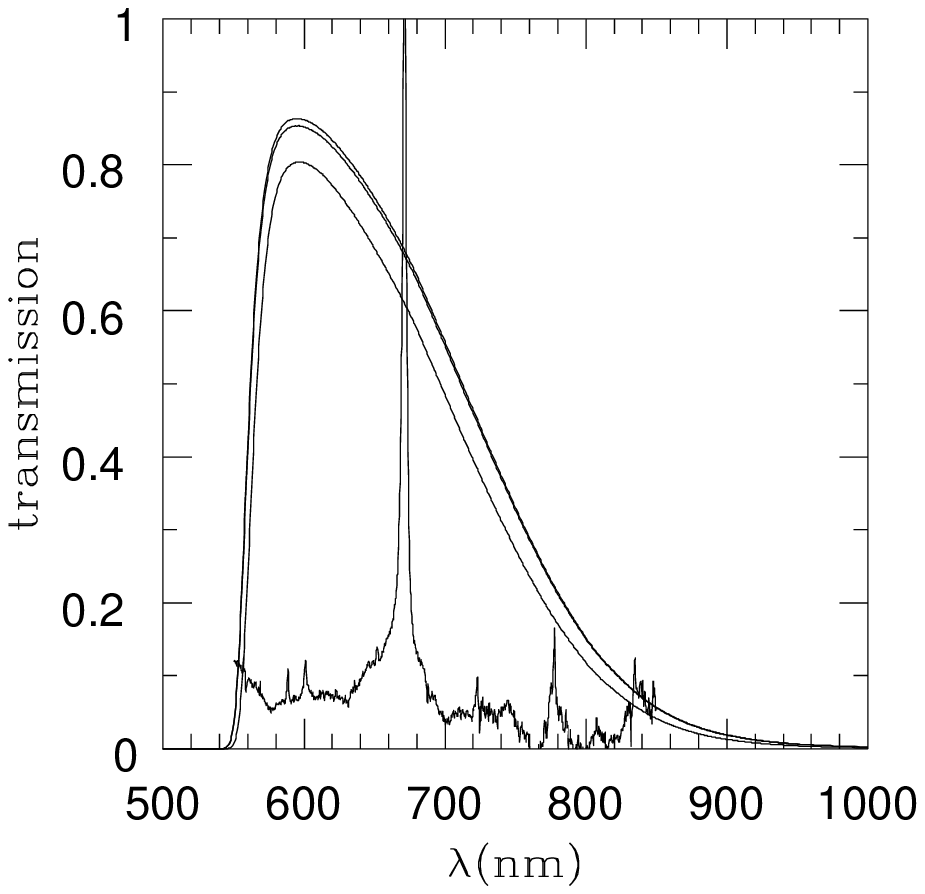}
   \caption{Transmission values of the three $R$-band filters used in
the observations (ESO\#554, ESO\#608, ESO\#642). Also plotted is the \Ha\
spectral region of SN~1988Z on 1989 April 5}
\end{figure*}


\ifoldfss
  \section{Spectral line analysis}
\else
  \section[]{Spectral line analysis}
\fi

Figure~4 and 5 show the evolution of the \Ha\ and \Hb\ profiles as a
function of age, from days 155 to 2981.

\begin{figure*}
   \cidfig{5in}{20}{144}{580}{750}{\DIRFIGS 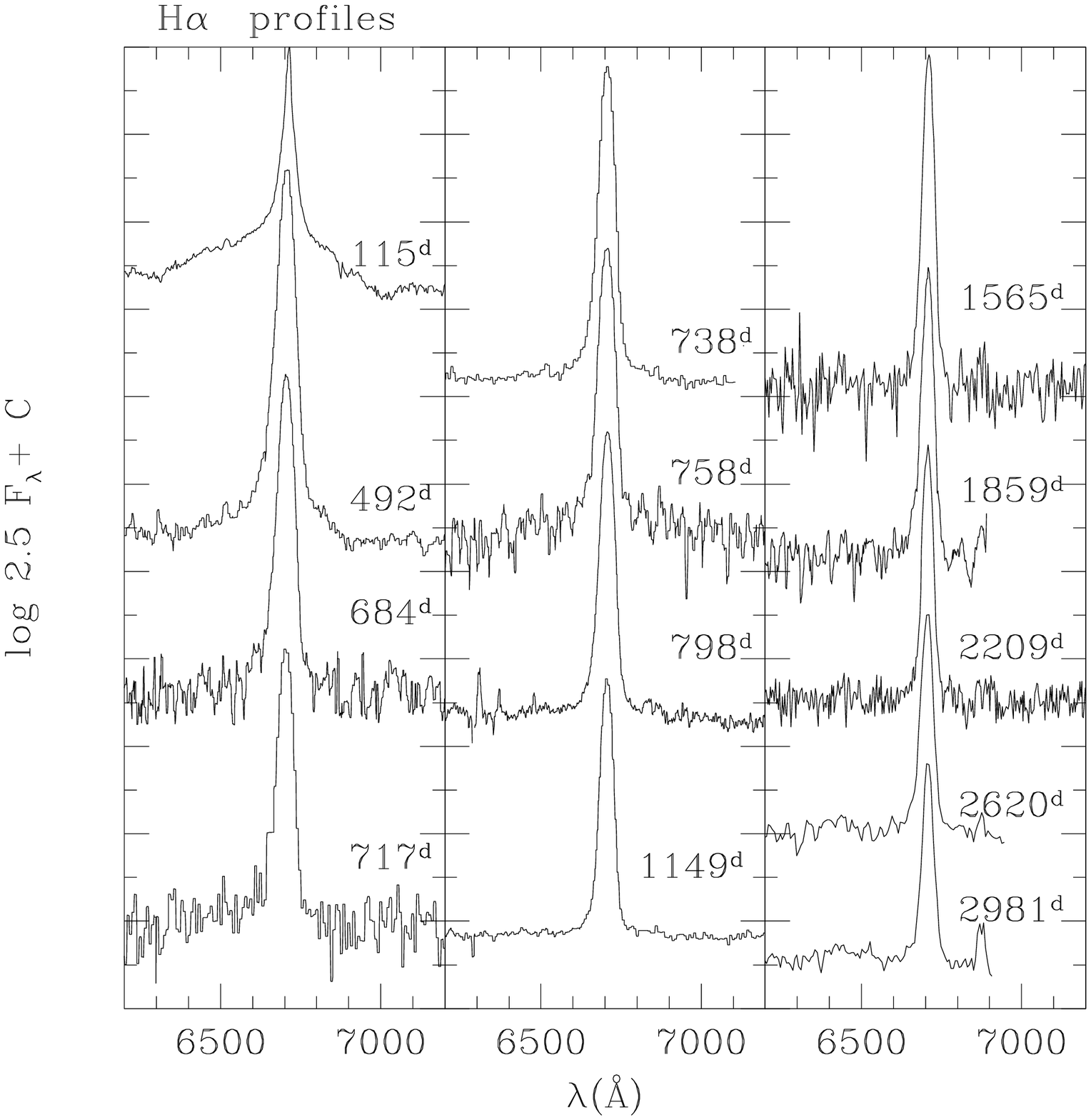}
   \caption{Evolution of \Ha\ profile}
\end{figure*}

\begin{figure*}
   \cidfig{5in}{20}{144}{580}{750}{\DIRFIGS 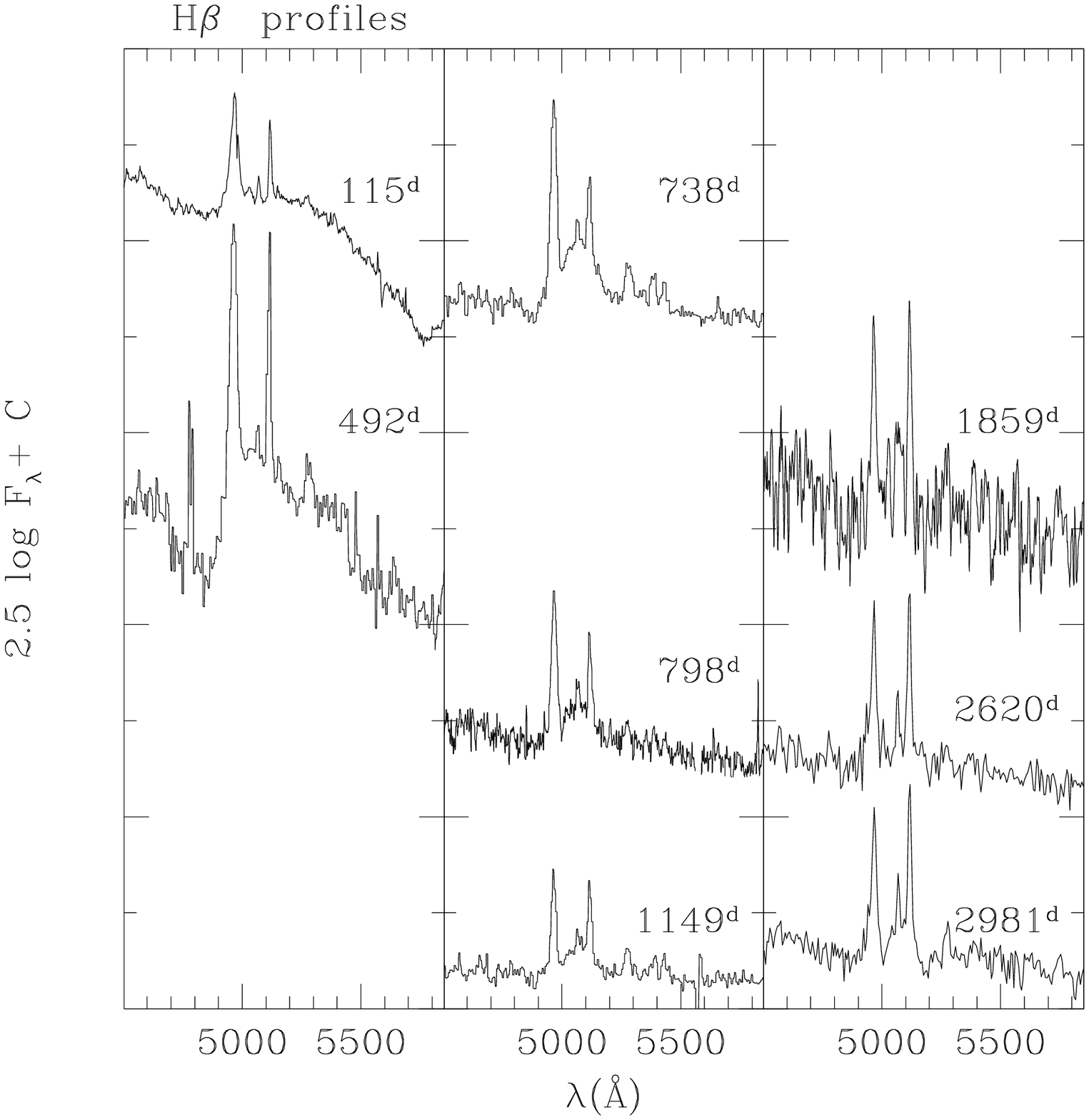}
   \caption{Evolution of \Hb\ profile}
\end{figure*}

Following the analysis performed by Turatto et al. (1993a),
we report in table~\ref{lines} the main emission line parameters as
measured with the {\sc ALICE} package in {\sc MIDAS}. 
This package allows multiple
gaussian fitting of complex line profiles. The coding for broad, 
intermediate and narrow components follows that adopted in Turatto et al. 
(1993a). These were the components identified
in the early evolution of SN~1988Z which we are following at later epochs. 
The analysis presented here is thus fully complementary 
with the results presented in that
report. Figure~6 shows the decomposition performed in the \Ha\ line
of the spectrum taken on 1996 February 14 as an example. In table~\ref{lines} 
 a colon indicates that the value is uncertain.

\begin{figure*}
   \cidfig{3in}{10}{156}{280}{440}{\DIRFIGS 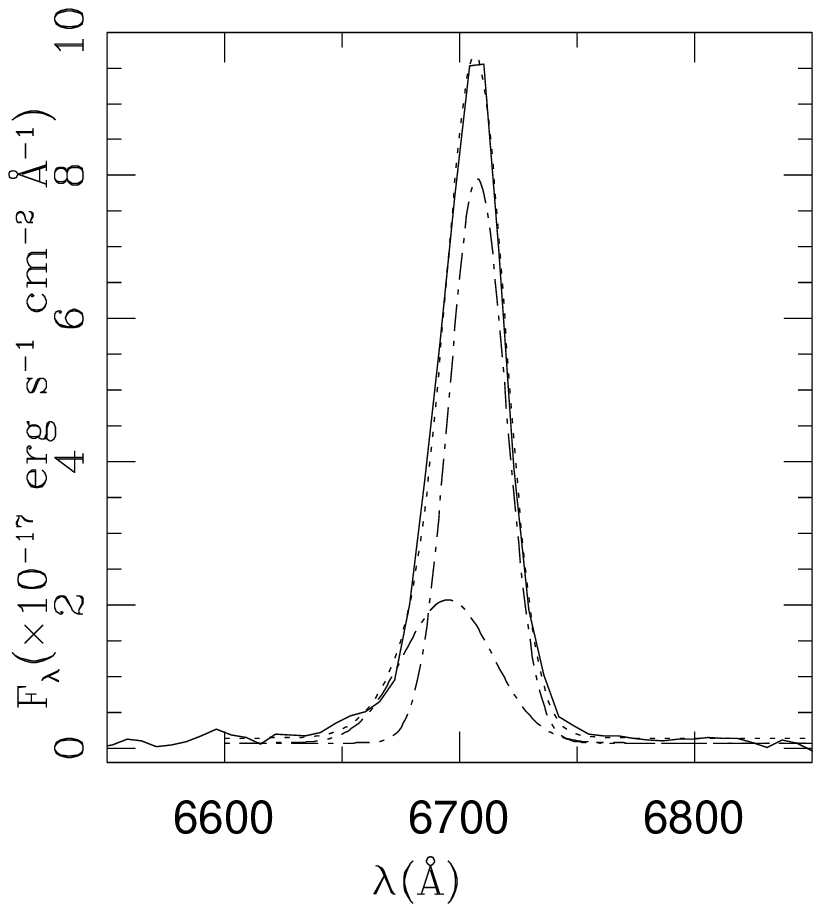}
   \caption{Multicomponent fit of the \Ha\ line of the spectrum taken on 
1996 February 14. The parameters of the components are given in table 5.
The dot-dashed lines represent single gaussian components, while the solid
line and dotted line represent the spectrum and the fit.}
\end{figure*}

The decrease of the width of the broad component of \Ha, shown in
figure~7 with filled squares, is
monotonical.
The behaviour can be described by a power-law, which is similar to
the time evolution of the velocity of the outer 
shock generated by a supernova remnant which evolves in a high-density
medium, $t^{-5/7}$ (eq. A5),  and 
which is plotted as a dashed-line.          
The law has been scaled to
the velocity
width of the line $\sim$ 4000 km/s at day 1000.
By comparison, the width of the intermediate component of \Ha\
(plotted with empty squares) decreases in a much slower way.

Figures~8 and 9 show the observed trend of the Balmer decrement and the
[O~III]\ldo{5007} luminosity with time. Both show an initial increase
followed by
a sharp down turn and a slower decay. Also shown in Figure~8 
are the predictions
of the simple model of cSNR evolution. 
In qualitative terms the observations follow
the trend prescribed by the simple model.
The discrepancies, though, strongly suggests that there are
fundamental aspects that are not considered by this  model.
These discrepancies together with the evolution of the line profiles
an multifrequency light curve, will be the basis for the construction
of more detailed numerical models which will include  
the role played by the density profile of the ejecta 
and the time dependent aspects
of the CSM evolution.

\begin{figure*}
   \cidfig{3in}{10}{125}{280}{440}{\DIRFIGS 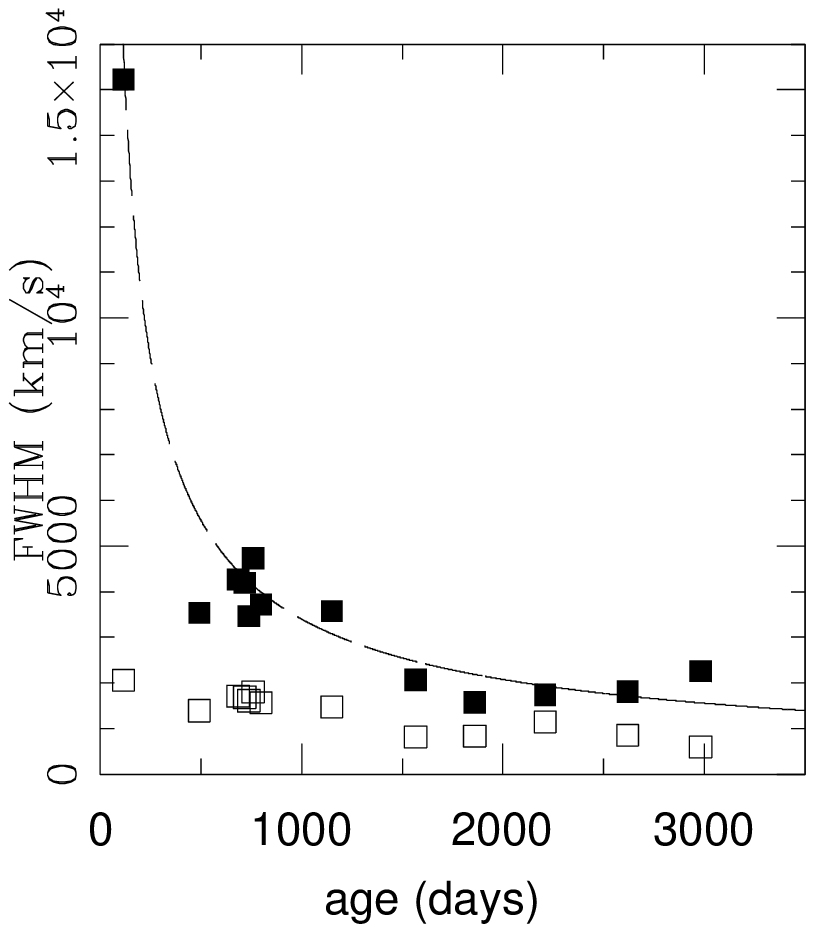}
   \caption{Evolution of the \Ha\ line profile. Solid squares represent 
the broad component and empty squares represent the intermediate component.
The dashed line represents the law for the velocity shock evolution of a cSNR.}
\end{figure*}

\begin{figure*}
   \cidfig{3in}{282}{160}{580}{435}{\DIRFIGS 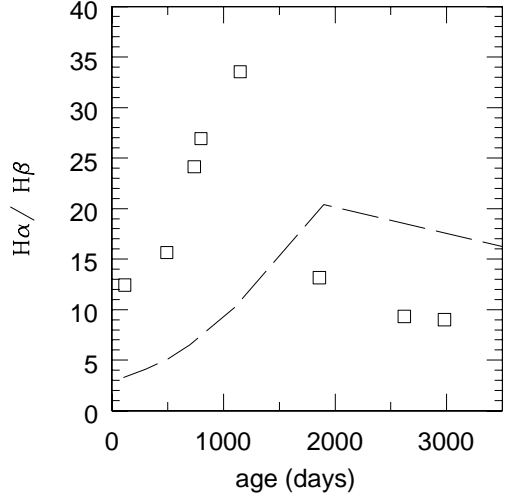}
   \caption{Evolution of the Balmer decrement (empty squares). The dashed 
line represents the theoretical values for the evolution of a cSNR.}
\end{figure*}

\begin{figure*}
   \cidfig{3in}{282}{160}{580}{435}{\DIRFIGS 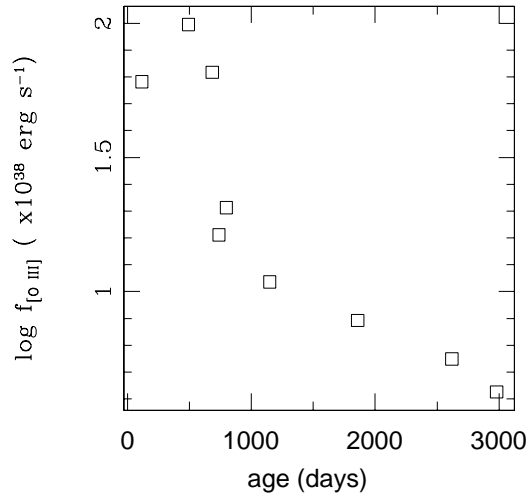}
   \caption{Evolution of [O III]\ldo{5007} line}
\end{figure*}

\begin{table*}
\begin{minipage}{140mm}
\begin{center}
\caption{Spectral line decomposition of the late spectra of SN~1988Z}
\label{lines}
\begin{tabular}{cllccccccccc}
\hline
\hline
 age & & & &H$\alpha$ &&&& &H$\beta$ & & HeI$-$NaID \\
\cline{4-6}\cline{8-11}
days && &broad& &interm.&&&interm.& &narrow&\\
\hline
1565&& FWHM (\AA)&48.6&&25.1&&& && &75:\\
       & & flux ($\times 10^{-16}$ erg\,s$^{-1}$\,cm$^{-2}$) &26&&98&&& &&
&6:\\
\hline
1859&& FWHM (\AA)&40.7&&28.3&&&60.9&&16.3&25.3\\
       & &  flux ($\times 10^{-16}$ erg\,s$^{-1}$\,cm$^{-2}$)
&8&&67&&&2.7&&3.0&2.2\\
\hline
2209&& FWHM (\AA)&49.2&&27.3&&& &&18.6&47:\\
       & & flux ($\times 10^{-16}$ erg\,s$^{-1}$\,cm$^{-2}$) &5&&47&&&
&&2&1.4:\\
\hline
2620&& FWHM (\AA)&45.1&&28.6&&&61.1&&16&21:\\
       & & flux ($\times 10^{-16}$ erg\,s$^{-1}$\,cm$^{-2}$)
&9.6&&24&&&1.7&&1.9&0.4:\\
\hline
2981&& FWHM (\AA)&53.9&&25.3&&&45.7&&16.7&51.2\\
       & & flux ($\times 10^{-16}$ erg\,s$^{-1}$\,cm$^{-2}$)
&6.9&&12.9&&&1.2&&1.3&1.0\\
\hline
\end{tabular}
\end{center}
\end{minipage}
\end{table*}


\ifoldfss
  \section{Photometric Analysis}
\else
  \section[]{Photometric Analysis}
\fi

\subsection{Spectral Energy Distribution}

 In figure~2 we observe that the light evolution in different bands 
peaks at different times: first in the optical, before discovery; 
then in \Ha\ (i.e. in the ionizing continuum), between 100 and 500~days; 
and lastly at radio-frequencies between 400 and 2000 days.
The spectral energy distribution (SED) thus changes with time.

In order to include the ionizing radiation, which is absorbed by the
cool material, 
we estimate its value from the flux of a prominent 
recombination line. \Ha\ is particularly useful, since several percent
of the ionizing energy is re-radiated through this line.
Assuming case~B recombination at 10000~K, which is the average temperature 
of the gas that emits broad lines in the hydrodynamical models of
cSNR (Terlevich et al. 1992), we can convert the \Ha\ flux to ionizing
flux. 
The ionizing flux $f_{\mbox{\scriptsize ion}}$ is related to the flux
of ionizing photons $q( \mbox{H}^0)$ by 
\begin{equation}
	f_{\mbox{\scriptsize ion}} = q( \mbox{H}^0) h 
\overline{\nu_{\mbox{\scriptsize ion}}}  \mbox{\ \ \ ,}
\end{equation}
where $\overline{\nu_{\mbox{\scriptsize ion}}}$ is the mean frequency
of the ionizing photons, approximately 27 eV. 
For case~B recombination, the \Ha\ flux is 
related to the flux of ionizing photons
by $f_{\Ha} = 1.367088 \times 10^{-12}q( \mbox{H}^0)$ in c.g.s. units 
(Kennicutt 1988, Osterbrock 1989).
Thus the conversion between the fluxes is 
\begin{equation}
	f_{\mbox{\scriptsize ion}} \approx 32 f_{\Ha}  \mbox{\ \ \ .}
\end{equation}

Figure~10 shows the SED at six stages in the evolution
of SN~1988Z. Solid squares correspond to the flux values which were 
interpolated
from those in table~4, and in the case of the ionizing continuum,
estimated using equation~3. 
Open squares indicate that we have performed
an extrapolation using the fits of Van Dyk et al. (1993) for 
radio wavelengths, or a $t^{-11/7}$ law drawn from the nearest value of
optical points and from the 
\Ha\
or X-rays for the hard radiation. Therefore, open symbols should be 
taken with caution. The only exception being values 
extrapolated from data which are $\pm50$~days away, in which case these
are represented as solid squares too.
Crosses correspond to the \Ha\ line-free $R_{\mbox{\scriptsize cont}}$ flux.

The SED of SN~1988Z differs from the SED of classical SNe in that its
high energy distribution is very prominent. There is also an indication
that the shape in the early evolution differs from the general
assumption that the emergent spectrum 
is created by an expanding hot
atmosphere with mainly a Planckian output (e.g. Catchpole et al. 1989,
Eastman \& Kirshner 1989). 
The dashed line in Figure~10 shows what the SED would look line if the
optical colours of SN~1988Z were to be interpreted as a blackbody
distribution, as has been assumed in the case in other type~IIn SNe 
(e.g. Wegner 
\& Swanson 1996). 
Hard-radiation and radio excesses are expected when the SN shock
interacts with a CSM.
At this stage we are viewing the kinetical reprocessing of a SN remnant

\subsection{Radiated energy estimates}

In order to calculate the energy radiated by SN~1988Z in the first eight
years of its evolution, first we integrate the light curves of Figure~2,
interpolating linearly in the time-axis between the observed points, 
i.e. with no further extrapolation on the behavior of the light curves
after the
last observational point. This will give us an estimate of the
observed radiated energy during these eight years, which should be
regarded as a lower limit to the total energy radiated during that
time. Table~6 gathers
the results,
where time refers to the maximum span of the light curve in that
band. From this table we derive that
the observed emitted energy of SN~1988Z is dominated by the optical
to X-ray radiation, having emitted at least 
$1.0\times10^{51}$~erg
in eight years.

This is a very rough estimate, which leaves gaps in
the spectral energy distribution and in the time evolution at
the different bands. As an example, if we interpolate linearly the
monocromatic flux
at frequencies between $B$ and $V$ and $V$ and $R$-bands along the
evolution
of the light curve, i.e.
\begin{equation}
	E_{\mbox{\scriptsize opt}} = 4 \pi d^2 \int_0^{8.2
	\mbox{\scriptsize yr}} dt \int^{\nu_B}_{\nu_R} d\nu f_\nu
\mbox{\ \ \ ,}
\end{equation}
where 
$d$ is the distance to the cSNR, then
we obtain that the energy radiated from $B$ to $R$ band is
$1.6\times 10^{50}$~erg, or $1.5\times 10^{50}$~erg if using 
$R_{\mbox{\scriptsize cont}}$
instead of $R$.

In order to obtain a more complete estimate of
the emission at optical-UV wavelengths, we have performed 
an extrapolation of the $B$, $V$ and $R_{\mbox{\scriptsize cont}}$ 
fluxes to cover the 
912\AA\ to $1\mu$m range, assuming that the SED is a
power-law, i.e. $f_\nu \propto \nu^{\alpha}$, where $\alpha$ is
calculated from the $B-V$ colours for the UV to $V$-band
interval  (typically $-3.0 < \alpha < -0.5$)
and from  the $V-R_{\mbox{\scriptsize cont}}$ 
colours for the $V$-band to 1$\mu$m
interval (typically $-2< \alpha < 5$). If the color doesn't change
dramatically
between these regions, the optical energy inferred is of the order of 
$3.2\times 10^{50}$~erg for the eight year evolution. 

  The ionization energy, in turn, can be estimated from the 
integration of the \Ha\ light curve. The use of eq.~3 gives a value of 
$8.7 \times 10^{50}$erg.

To estimate the radiation at 0.2--2~keV, we have extrapolated 
the three X-ray points to earlier epochs than 6.5~yr, when the first
ROSAT observation was obtained. 
If we
adopt the $t^{-11/7}$ law before day 2355, so that at
day 2355 it reproduces the first ROSAT observation, we have that
the energy radiated in 3000 days in the 0.2--2~keV band 
could be as high as $6.0\times10^{50}$~erg. 
If a complete bremsstrahlung spectrum of 1 keV 
is considered, the energy emitted in
X-rays
alone could be about $9.6\times10^{50}$~erg, and if we consider internal
absorption of the CSM, this could go up to $7.6\times10^{51}$.

 These estimated values of the emitted energy in the 8.5~yr of 
evolution 
of SN~1988Z at different wavelengths are summarized in Table~7.
Beware that some entries in this table are subsets of other entries
(e.g. 4000--7000\AA\ and 912--1$\mu$m).
and that some entries are in part the result of reprocessing the energies 
of other entries (e.g. 912--1$\mu$m and bremsstrahlung).
Considering that half of the leading shock radiated energy is emitted
outwards and the other half inwards and is therefore 
reprocessed by the
high density thin shell, the total energy radiated in the event
should include either twice the observed X-ray luminosity or
the sum of the observed X-ray luminosity and the inferred
ionizing and optical luminosity, or twice the sum of the observed
ionizing and optical luminosity.

A strong lower limit of the total radiated energy is then
$2\times 8.7\times10^{50} + 3.2\times10^{50} \approx
2 \times10^{51}$~erg. 
The bolometric light curve derived for this estimate is represented with 
a thick solid line in figure~2.

A best estimate of the total radiated energy requires a 
model-dependent extrapolation either from the X-ray or from the \Ha\
radiation. The X-ray derived values are rather uncertain, 
since they require the assumption of time, spectral and absorption
corrections. Allowing for these uncertainties, the 
energy could be as high as $2 \times 7.6\times10^{51} \approx
10^{52}$~erg.
The bolometric light curve derived for this model-dependent estimate is 
represented with a thick dashed line in figure~2. Note that the excess
energy of this second estimates is produced mainly at early epochs.
A high value of the X-ray radiated energy is supported by the 
ionization energy estimates. The models indicate that during
the earlier stages of evolution
the emitted radiation is very hard and most of it leaves the
remnant without producing ionization.
A rough estimate based on the simple model indicates that
the total energy radiated is about six times larger than the
ionizing radiation, that is about $4\times10^{51}$. Recalling that
this includes only half of the total energy emitted, we 
conclude again that the total energy radiated may be close to $10^{52}$.

We emphasize that these estimates 
totally disregard any intrinsic reddening produced in the host 
galaxy or in the CSM of SN~1988Z.

 \begin{table}
 \begin{center}
  \caption{Integrated values of observed emitted energy 
in different wavelengths}
  \begin{tabular}{rcc}
\hline
band & time  & energy   \\
     & yr    & erg    \\	
\hline
0.2--2 keV & 2.0 &  $3.6\times10^{48}$\\
ionization    & 8.2   &	$8.7\times10^{50}$\\
$B$    & 4.3   &	$4.0\times10^{49}$\\
$V$    & 6.1   &	$2.6\times10^{49}$\\
$R$    & 8.2   &	$5.8\times10^{49}$\\
$R_{\mbox{\scriptsize cont}}$   & 8.2   &	$3.7\times10^{49}$\\
2 cm  & 2.3   &  $2.4\times10^{46}$\\
3.6 cm& 2.4   &  $2.2\times10^{46}$\\
6 cm  & 3.0   &  $1.4\times10^{46}$\\
20 cm & 2.3   &  $1.4\times10^{45}$\\
\hline
\end{tabular}
\end{center}
\end{table}

 \begin{table}
 \begin{center}
  \caption{Integrated values of estimated emitted energy 
in different wavelengths during the 8.5 yr of evolution}
  \begin{tabular}{rl}
\hline
band & energy    \\
     & erg   \\	
\hline
bremsstrahlung	 &   $9.6\times10^{50}$ to $7.6\times10^{51}$\\
0.2--2 keV       &   $6.0\times10^{50}$ to $4.8\times10^{51}$\\
ionization              &   $8.7\times10^{50}$  \\
4000--7000\AA    &   $1.5\times10^{50}$ \\
912\AA -- 1$\mu$m &   $3.2\times10^{50}$ \\
\hline
\end{tabular}
\end{center}
\end{table}

\begin{figure*}
   \cidfig{5in}{20}{144}{580}{710}{\DIRFIGS 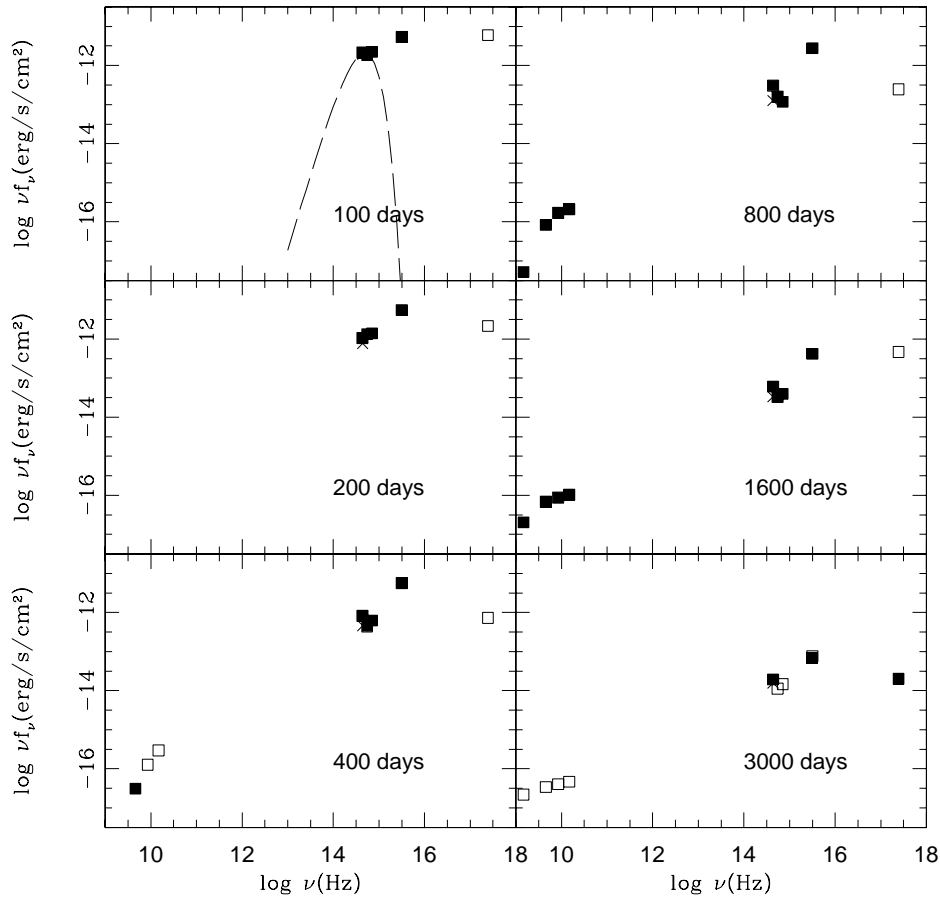}
   \caption{SED evolution of SN~1988Z:
solid squares correspond to interpolations from values actually observed, 
and empty squares correspond to extrapolations. Crosses correspond to the 
\Ha\ line-free $R_{\mbox{\scriptsize cont}}$ 
flux. The dashed line represents a blackbody at 5800~K.
}
\end{figure*}

\section{Discussion}

We have shown that the integrated light emitted in 
radio, optical and X-rays by SN~1988Z in the 8.5~yr of evolution after 
discovery  is of the order of several times $10^{51}$~erg and may be close 
to $10^{52}$~erg,
i.e. larger than  the canonical value of the kinetic energy released
in a SN explosion, while typical SNe radiate one to two orders of magnitude
less energy (e.g. Panagia et al. 1980, Catchpole et al. 1987, 1989).
The value of the radiated energy in SN~1988Z is slightly smaller
than that estimated for the most energetic kinetical release in a SN 
explosion observed so far, $(2-5)\times10^{52}$~erg, 
in SN~1998bw (Galama et al. 1998, Iwamoto et al. 1998).
Indeed on the basis of our result we
speculate that SN 1988Z  was a hypernova and may have been caused by the
formation of a black hole in the core of a massive star.
Most of the energy in SN~1988Z is radiated
in the optical to X-ray interval, with spectra, SED and light curves 
departing from 
those of classical SN~II. The high densities derived 
from the O forbidden lines and the flat light-curves strongly support 
 the idea of a
high-density circumstellar shell reprocessing the kinetic energy through fast
radiative shocks. 

The estimated CSM mass swept by the shock in this time
is of the order of 15\Msun\ (eq. A8), although this quantity is dependent on 
the assumed model. Chugai \& Danziger (1994), for example, give an estimate
of only about 1\Msun\ from their own models.

While a detailed comparison of light-curve, SED, and 
line profile evolution is still missing,
we have shown that the simplified model of ejecta interacting
with a dense and homogeneous CSM outlined in the appendix reproduces the
overall decay of the optical light, the level of X-ray and \Ha\ emission
and its line-width evolution. The evolution of the \Ha\ flux is however 
more rapid than that predicted by the canonical models, and the 
absolute values of line-widths and Balmer decrements differ sometimes 
by more than a factor of 2. More
fine-tuned models, possibly introducing a small 
density gradient in the circumstellar shell,  are thus required.

Observationally much work is also needed, on one hand to identify cSNR
like SN~1988Z early
enough in their evolution, and on the other to obtain good quality
data over the widest possible spectral range and time coverage.
High-quality and high-resolution spectral information is particularly 
needed in the X-ray region early enough in the evolution of the SN, 
where and when most of the energy is radiated. 
A  close optical monitoring of these objects would allow us to assess
the predicted existence of modulations in the light 
curve decay and line component anomalities in the spectra 
due to catastrophic radiative cooling,
and the precise dating of the 
disappearance of the broad emission lines.

\section*{Acknowledgments}
We would like to thank G. Tenorio-Tagle for useful suggestions on an
early version of this paper.
This work has been supported in part by the `Formation and Evolution of
Galaxies' network set up by the European Commission under contract 
ERB FMRX-CT96-086 of its TMR programme.

\appendix

\section[]{The ``Simple model'' for compact supernova remnants}

The interaction of the SN ejecta with a dense CSM
causes a shocked region of hot gas enclosed by two shock waves: on the
outside the leading shock, and on the inside the inward facing
``reverse" shock. The leading shock ($v_s \sim 10^4$~\univel) encounters
dense circumstellar material and raises its temperature to $\sim 10^9$
\kelvin. The reverse shock, which is initially substantially slower ($v
\sim 10^3$~\univel), begins to thermalize the supernova ejecta to
temperatures of about $10^7$\kelvin . Early analytical and numerical
computations of the evolution of SNRs in a dense medium (Chevalier
1974; Shull 1980; Wheeler et al. 1980) showed a speeded-up evolution
compared with the ``standard" solution in a medium of $n_0$ = 1 \percucm.
All evolutionary phases (free expansion, thermalization of the
ejecta, the quasi-adiabatic Sedov phase, the radiative and the pressure
modified snow-plough phases) which have been thoroughly studied for the
standard case, are substantially speeded up as a consequence.

Supernova remnants evolving in a dense  CSM
 reach their maximum luminosity ($L > 10^8$~\Lsun) at small radii
($R <
0.1$~pc ) soon after the SN explosion ($t < 20$~yr) while still expanding
at
velocities of more than $1000$~\univel (Shull 1980; Wheeler et al.  1980;
Draine
and Woods 1991; Terlevich et al. 1992; Terlevich 1994).  In these cSNR,
radiative cooling becomes important well before the thermalization of the
ejecta is complete. As a result, the Sedov phase is avoided and the
remnant
goes directly from the free expansion phase to the radiative phase. The 
shocked matter undergoes a rapid condensation behind both the leading and
the
reverse shocks. As a consequence two  high-density, fast-moving thin
shells are
formed. These dense shells, the freely expanding ejecta and a section of
the
still dynamically unperturbed interstellar gas, are all ionized by the
radiation from the shocks.  

For SNe evolving in CSM densities of the order of
$n_0 \sim 10^7$ cm$^{-3}$, the cooling-time ($t_c$) and cooling-length
($r_c$)
scales for the post-shock temperatures are very small.

\begin{equation}
{ t_c \simeq {3kT_s\over 8n_0\Lambda} \simeq 0.2{v_8^2\over
n_7\Lambda_{23}}\ \ {\rm yr},}
\end{equation}

\begin{equation}
{ r_c \simeq {1\over 4}t_cv_s \simeq 1.8\times 10^{14}
{v_8^3\over n_7\Lambda_{23}}\ \ {\rm cm}, }\
\end{equation}

\noindent
where $v_8$ is the velocity in $10^8$~km~s$^{-1}$ units, 
$n_7$ is the density in  $10^7$\percucm units, 
$\Lambda_{23}$ is the cooling function ($\Lambda$) in 
$10^{-23}$ erg cm$^3$ s$^{-1}$ units,  
$T_s$ is the shock temperature, and $k$ is Boltzmann's constant. 
Thus, radiative losses become important at very early
times when the shock velocities and temperatures are
$v_s>10^3$~km~s$^{-1}$
and $T_s>10^7$ K, well before the ejecta is even thermalized.
This means that a large flux of ionizing photons will emerge from the
shocked gas at X-ray energies.

For a supernova remnant which injects $10^{51}$ ergs into a medium of
constant density $\n7$, the onset of the
radiative phase behind the leading shock 
which causes the formation of a dense outer shell, assuming 
free-free cooling only, (Shull 1980, Wheeler
et al.  1980, Draine and Woods 1991) begins at a time $t_{sg}$, given by

\begin{equation}
{t_{sg} = 230\;E_{51}^{1/8}n_7^{-3/4} \mbox{ \ \ days} }
\end{equation}

\noindent
where $E_{51}$ is the energy deposited by the SN in units of
$10^{51}$~ergs.
At this stage, the shock is at a radius 
\begin{equation}
{R_{s} = 0.01\;E_{51}^{1/4}\;n_7^{-1/2}\;
\Bigl({t\over t_{sg}}\Bigr)^{ 2/7}\; \mbox{pc,}  }
\end{equation}
with velocity
\begin{equation}
{v_{s} = 4600 \;E_{51}^{1/8}\;n_7^{1/4} \;
\Bigl({t\over t_{sg}}\Bigr)^{ -5/7}\univel } \mbox{\ \ ,}
\end{equation}
temperature
\begin{equation}
{T_{s} = 3.0\times 10^8 \;E_{51}^{1/4}\;n_7^{1/2}\;\Bigl({t\over
t_{sg}}\Bigr)^{-10/7}\; \mbox{K} } \mbox{\ \ ,}
\end{equation}
and luminosity
\begin{equation}
{L_{s} = 2\times 10^{43}\;E_{51}^{7/8}\;n_7^{3/4}\;
\Bigl({t\over t_{sg}}\Bigr)^{-11/7}\;\ergsec. }
\end{equation}
The mass swept by the shock is
\begin{equation}
M_{shell} = 1.1 \;E_{51}^{3/4}\;n_7^{-1/2}\;\Bigl({t\over
t_{sg}}\Bigr)^{6/7}\;\Msun  \mbox{.}
\end{equation}

These approximate formulae assume that the ejecta has already been
fully thermalized. However, for the case of interest in this work (i.e.
for $n_0$ $\geq$ $10^{5}$ \percucm) strong radiative losses occur
before thermalization is complete. Because the cooling processes
radiate the thermal energy at the same rate as thermalization proceeds,
the Sedov phase is totally inhibited and thus there is no
self-consistent analytic treatment for the evolution of such remnants.
It is therefore necessary to follow the detailed time evolution of the
gas flow. In particular, special care should be given to the post-shock
structure which is sensitive to the details of the ambient density
distribution and to the temperature dependence of the radiative cooling
function.

The cooling time scale, $t_c$, for an optically thin plasma, is
proportional to the inverse of the gas density and the evolution
proceeds faster at higher ambient densities. On the other hand, the
temperature dependence of the cooling function is different for
different temperature ranges. Adiabatic shocks are stable but cooling
instabilities can develop over a wide range of radiative shock
conditions (Avedisova 1974, Falle 1975, 1981, McCray, Stein and Kafatos
1975, Chevalier and Imamura 1982, Imamura 1985, Bertschinger 1986).

The details of the transition from a nearly adiabatic to a strongly
radiative shock depend on the ability of the gas to readjust to the
cooling rate. Pressure gradients tend to be smoothed out in a
sound-crossing time, $t_d$, and the ratio $t_c/t_d$ provides an
estimate of the conditions prevailing in the cooling gas. For
$t_c/t_d>1$, at moderate cooling rates, the gas elements are
continuously compressed as their temperature falls and the cooling
process operates quasi-isobarically at the pressure attained by the gas
immediately behind the shock. For $t_c/t_d<1$, however, the cooling
rate dominates over any pressure readjustment and the process becomes
quasi-isochoric at the post-shocked density of the cooling gas elements.

A large pressure imbalance then develops in the flow, and new
additional shocks are generated which end up compressing the cooled
gas. This process, termed ``catastrophic cooling'' (Falle 1975, 1981),
appears during thin shell formation and the instabilities continue to
operate during the rest of the radiative shock evolution (Chevalier and
Imamura 1982; Bertschinger 1986; Cioffi, McKee \& Bertschinger 1988; 
Tenorio-Tagle et al
1990). The catastrophic cooling acquires a central role in the case of
supernovae evolving in high density media due to the strength of the
radiation produced upon cooling, and the rapid variations inherent in
the shock propagation. These features imply that a large flux of
ionizing photons will emerge from the shocked gas. The wide range of
gas temperatures in the cooling region results in a power-law-like
spectrum at UV and X-ray frequencies (Terlevich et al. 1992).


\begin{thebibliography}{99}

	\bibitem{key1}Avedisova, V. S., 1974, Sov Astr., {18}, {283}
 
     \bibitem{BCT}\rev{Barbon R., Cappellaro E. \& Turatto
M.}{1989}{AAS}{81}{421}
      \bibitem{BCR79}\rev{Barbon R., Ciatti F. \& Rosino L.}{1979}{AA}{72}{287}
	\bibitem{key2} Bertschinger E., 1986, {ApJ}, {304}, {154}
      \bibitem{BP92}\rev{Bregman J.N., Pildis R.A.}{1992}{ApJ}{398}{107}
      \bibitem{BH79}\rev{Burstein D. \& Heiles C.}{1982}{AJ}{87}{1165}

      \bibitem{C88} Cappellaro E. \& Turatto M. 1988, IAU Circ. 4691
	\bibitem{C87}\rev{Catchpole et al.}{1987}{MNRAS}{229}{15P}
	\bibitem{C89}\rev{Catchpole et al.}{1989}{MNRAS}{237}{55P}
	\bibitem{key6}Chevalier R. A., and Imamura J. N., 1982, 
{ApJ}, {261}, {543}
      \bibitem{key4}Chevalier R. A., 1974, {ApJ}, {188}, {501}
      \bibitem{C94}\rev{Chugai N.N.}{1991}{MNRAS}{250}{513}
      \bibitem{C94}\rev{Chugai N.N. \& Danziger}{1994}{MNRAS}{268}{173}
	\bibitem{key7}Cioffi D. F., McKee C. F., 
and Bertschinger, E., 1988, {ApJ}, {334}, {252}

\bibitem{key11}Draine B.T. and Woods D.T., 1991, {ApJ}, {383},
{621}

      \bibitem{FT96}\rev{Eastman R.G. \& Kirshner R.P.}{1989}{ApJ}{347}{777}

      \bibitem{FT96}\rev{Fabian A.C., Terlevich
R.J.}{1996}{MNRAS}{280}{L5}
	\bibitem{key12}Falle S. A. E. G., 1975, {MNRAS}, {172},{55}
	\bibitem{key13}Falle S. A. E. G., 1981, {MNRAS}, {95}, {1011}
	\bibitem{f90} Ferland G.J., 1997, {\it Hazy, a brief introduction
to CLOUDY 90'}, Univ. of Kentucky at Lexington. 
      \bibitem{F97}\rev{Filippenko, A.V.}{1997}{ARAA}{35}{309}
      \bibitem{F91} Filippenko A.V. 1991, in `Supernovae',
      Ed. S.E. Woosley, Springer-Verlag, page 467.
      \bibitem{F89}\rev{Filippenko A.V.}{1989}{AJ}{97}{726}

	\bibitem{G88} \rev{Galama T.J. et al.}{1998}{Nat}{395}{67}
      \bibitem{GK89}  Gaskell C.M., Koratkar A.P., 1989, IAU Circ. 4761

      \bibitem{H94}\rev{Hamuy M., Suntzeff N.B., Heathcote S.R., 
Walker A.R., Gigoux P., Phillips M.M.}{1994}{PASP}{106}{566}

	\bibitem{key16} Imamura J. N., 1985, {ApJ}, {296}, {128}
	\bibitem{I98} Iwamoto K. et al., 1998, Nat, 395, 672.
	\bibitem{k88} Kennicutt R.C., 1988, ApJ, 334, 144

      \bibitem{L92}\rev{Landolt A.U.}{1992}{AJ}{104}{340}

	\bibitem{key18}McCray R, Stein R. F., and Kafatos 
M., 1975, {ApJ}, {196}, {565}

     \bibitem{O89} Osterbrock D.E. 1989, ``Astrophysics of Gaseus
     Nebulae and Active Galactic Nuclei'', University Science Books.

      \bibitem{P80} Panagia N. et al. 1980, MNRAS, 192, 861.
	\bibitem{P95} Plewa T., 1995, MNRAS, 275, 143.
      \bibitem{P88} Pollas C., 1988a, IAU Circ. 4691
      \bibitem{P88b} Pollas C., 1988b, IAU Circ. 4696
      \bibitem{P89} Pollas C., 1989, IAU Circ. 4742

      \bibitem{S98}\rev{Salamanca I., Cid Fernandes R., Tenorio-Tagle G., 
Telles E., Terlevich R.J., Mu\~noz-Tu\~non C.}{1998}{MNRAS}{300}{L17}
      \bibitem{S90}\rev{Schlegel E.M.}{1990}{MNRAS}{244}{269}
      \bibitem{S90}\rev{Seaton M.J.}{1979}{MNRAS}{187}{73P}
      \bibitem{S90}\rev{Shull J.M.}{1980}{ApJ}{237}{769}	
      \bibitem{SS91}\rev{Stathakis R.A., Sadler
E.M.}{1991}{MNRAS}{250}{786}

	\bibitem{key24} Tenorio-Tagle G.,
Bodenheimer P., Franco J., and Rozyczka M., 1990, {MNRAS}, {244},
{563}
     \bibitem{Te94} Terlevich R.J., 1994, in Clegg R.E.S., Stevens I.R.,  Meikle W.P.S., eds, Circumstellar Media in the Late Stages of Stellar Evolution. Cambridge Univ. Press, Cambridge, p. 153
     \bibitem{T92} \rev{Terlevich R.J., Tenorio-Tagle G., Franco J.,
Melnick J.}{1992}{MNRAS}{255}{713}
	\bibitem{Y99} Turatto M., Benetti S., Cappellaro E., Danziger I.J.,
Mazali P.A., 1999, in {\it 'SN 1987A: Ten Years After'}, eds. M.M. 
Phillips, N.B. Suntzeff, Fifth CTIO/ESO/LCO Workshop, in press.
      \bibitem{T93}\rev{Turatto M., Cappellaro E., Danziger I.J.,
Benetti S., Gouiffes C., Della Valle M.}{1993a}{MNRAS}{262}{128}
	\bibitem{T93b} Turatto M., Cappellaro E., 
Benetti S., Danziger I.J., 1993b, MNRAS, 265, 471.

      \bibitem{vD93}\rev{Van Dyk S.D., Weiler K.W., Sramek R.A., Panagia N.}{1993}{ApJL}{419}{L69}

      \bibitem{WS96} Wegner G. \& Swanson S.R., 1996, MNRAS, 278, 22.
      \bibitem{S90}\rev{Wheeler J.C., Mazurek T.J. \&
Sivaramakrishnan}{1980}
{ApJ}{237}{78}


\end{thebibliography}
\end{document}